\documentclass{article}
\usepackage{spconf,amsmath,graphicx}

\usepackage{subfigure}
\usepackage{amsopn, amsthm, epsfig}
\usepackage{amsfonts}       

\usepackage{algorithm, algpseudocode}

\title{It\^oWave: It\^o Stochastic Differential Equation Is All You Need For Wave Generation}
\twoauthors
 {Shoule Wu\sthanks{Equal contribution.}}
	{Yangzhou University\\
	Yangzhou, China}
 {Ziqiang Shi\footnotemark[1]}
	{Fujitsu R\&D Center\\
	Beijing, China}
\begin{document}
%
\maketitle
\begin{abstract}
  In this paper, we propose a vocoder based on a pair of forward and 
  reverse-time linear stochastic differential equations (SDE). The solutions of this SDE pair are two stochastic 
  processes, one of which turns the distribution of 
  wave, that we want to generate, into a simple and tractable distribution. The other is the generation procedure 
  that turns this tractable 
  simple signal into the target wave. The model  is called It\^oWave. It\^oWave use the Wiener process as a driver 
  to gradually subtract the excess 
  signal from the noise signal to generate realistic corresponding meaningful audio respectively, under the conditional 
  inputs of original 
mel spectrogram. The results of the experiment show that the mean opinion scores (MOS) of   It\^oWave can exceed 
  the current state-of-the-art (SOTA) methods, and reached 4.35$\pm$0.115.
  The generated audio samples are available online\footnotemark[1].
\end{abstract}

\footnotetext[1]{https://wushoule.github.io/ItoAudio/.}

\section{Introduction}
\label{sec:intro}

The vocoder model is roughly categorized as autoregressive (AR) or non-autoregressive (non-AR), where the AR model generates the signal 
frame by frame, and the generation of the current signal frame depends on the previously generated signal.  Non-AR models generate the signal in parallel, and the 
current signal frame does not depend on the previous signal. Generally speaking, the voice quality generated by the AR model is 
higher than the non-AR model, but the amount of computation is also larger, and the generation speed is slow. While for the non-AR generation model, the generation 
speed is faster, 
but the generated voice quality is slightly worse. To name a few, for example, WaveNet~\cite{oord2016wavenet} is the earliest AR model, using sampling points as the unit and achieves a sound quality 
that matches the naturalness of human speech. In addition, other recent AR models, including sampleRNN~\cite{mehri2016samplernn} and LPCNet~\cite{valin2019lpcnet}
have further improved the sound quality. 
However, due to the large amount of 
computation and the slow generation speed, researchers currently mainly focus on developing non-AR wave generation models, such as Parallel WaveNet~\cite{oord2018parallel}, GanSynth~\cite{engel2019gansynth},  MelGan~\cite{kumar2019melgan}, WaveGlow~\cite{prenger2019waveglow}, 
Parallel WaveGan~\cite{yamamoto2020parallel}, and so on.

In this paper, vocoder is modeled with a new framework based on linear It\^o stochastic differential equations (SDE) and score matching modeling. 
We call it It\^oWave. The linear It\^o SDE, driven by the Wiener process, can slowly 
turn the  wave data distributions into data distributions that are easy to manipulate, such as white noise. 
This transformation process is the stochastic process solution of the linear It\^o SDE. Therefore, the corresponding  reverse-time linear 
It\^o SDE can generate the  wave data distribution required by vocoder,  from this easy data distribution, such as white noise. It can be seen that the
 reverse-time linear It\^o SDE is crucial for the generation, 
and Anderson~\cite{anderson1982reverse} shows the 
explicit form of this reverse-time linear SDE, and the formula shows that it depends on the gradient of the log value of the probability density function 
of the stochastic process solution of the forward-time equation. This gradient value is also called the stein score~\cite{hyvarinen2005estimation}. 
It\^oWave predict the stein score corresponding to the  wave by trained neural networks. 
After obtaining this score, It\^oWave can achieve 
the goal of generating wave through reverse-time linear It\^o SDE or Langevin dynamic sampling.

Our contribution is as follows, 1) We are the first to proposed a vocoder model based on linear It\^o SDE, and
 reached state-of-the-art performance; 2) We explicitly put vocoder  under a more flexible framework, which can construct different 
vocoder models by selecting different  drift and diffusion coefficients of the linear SDE;
3) For It\^oWave, we propose a network structure, which is suitable for estimating the gradient of log value of 
the density function of the 
wave data distributions.


\section{The It\^oWave}

\subsection{Audio data distribution transformation based on It\^o SDE}
\label{sec:dist_transormation_sde}

It\^o SDE is a very natural model that can realize the transformation between different data distributions. The general 
It\^o SDE is as follows
\begin{align}
  \left\{
  \begin{array}{lr}
 d\mathbf{X}=\mathbf{f}(\mathbf{X},t)dt+g(t)d\mathbf{W} &  \\
 \mathbf{X}(0)=\mathbf{x}(0) &
\end{array}
\right.
 \label{eq:ito_sde}
\end{align}
for $0\leq t\leq T$, where  $\mathbf{f}(\cdot ,t)$ is the drift coefficient,
 $g(t)$ is the diffusion coefficient,
 $\mathbf{W}$ is the standard Wiener process. Let $p(\mathbf{x}(t))$ be the density of the random variable $\mathbf{X}(t)$. 
This SDE~(\ref{eq:ito_sde}) changes the initial distribution $p(\mathbf{x}(0))$ into another distribution 
$p(\mathbf{x}(T))$
by gradually adding the noise from the Wiener process $\mathbf{W}$.
In this work, $\mathbf{x}(t)\in\mathbb{R}^d$, and $p(\mathbf{x}(0))$ is to denote the data distribution of wave in It\^oWave. 
$p(\mathbf{x}(T))$ is an easy tractable distribution (e.g. Gaussian) of the latent representation of the wave signal 
corresponding to the conditional 
mel spectrogram.
If this stochastic process $\mathbf{x}(t)$  can be reversed in time, then the corresponding target waveform of the conditional mel spectrogram can be generated
  from a simple latent distribution. 

Actually the reverse-time diffusion process is the solution  of the following corresponding reverse-time It\^o SDE~\cite{anderson1982reverse}
\begin{align}
  \left\{
  \begin{array}{lr}
 d\mathbf{X}=\left[\mathbf{f}(\mathbf{X},t)-g(t)^2\nabla_\mathbf{x}\log p(\mathbf{x}(t))\right]dt
 &  \\ \qquad \qquad +g(t)d\overline{\mathbf{W}}  &  \\
 \mathbf{X}(T)=\mathbf{x}(T) &
\end{array}
\right.
 \label{eq:reverse_ito_sde}
\end{align}
for $0\leq t\leq T$,
where $p(\mathbf{x}(t))$ is  the distribution of $\mathbf{X}(t)$, 
$\overline{\mathbf{W}}$ is the standard Wiener process in reverse-time.
The solution of this reverse-time It\^o SDE~(\ref{eq:reverse_ito_sde}) can be used to generate 
wave data from a tractable  latent distribution $p(\mathbf{x}(T))$. 
Therefore, it can be seen from~(\ref{eq:reverse_ito_sde}) that the key to generating wave with SDE lies in the calculations of 
$\nabla_\mathbf{x}\log p(\mathbf{x}(t))$ $(0\leq t\leq T)$, which is always called score function~\cite{hyvarinen2005estimation,song2019generative} of 
the data.

\subsection{Score estimation of audio data distribution}
\label{sec:score_estimation}


In this work, a neural network $\mathfrak{S}_{\theta}$ is used to approximate the score function, where $\theta$ denotes the parameters of the network. 
The input of the network $\mathfrak{S}_{\theta}$  includes time $t$, $\mathbf{x}(t)$, and conditional input  mel spectrograms $\mathbf{m}$ input in It\^oWave.
The expected output 
is $\nabla_{\mathbf{x}(t)} \log  p(\mathbf{x}(t))$.  
The objective of score matching is~\cite{hyvarinen2005estimation} 
\begin{align}
  \mathbb{E}_{t\sim [0,T]}\mathbb{E}_{\mathbf{x}(t)\sim p(\mathbf{x}(t))} 
 [ \frac{1}{2}\parallel \mathfrak{S}_{\theta}(\mathbf{x}(t),t,\mathbf{m}) \nonumber \\ -\nabla_{\mathbf{x}(t)}  \log  p(\mathbf{x}(t))\parallel^2].
  \label{eq:esm_loss}
\end{align}

Generally speaking, in the low-density data manifold area, the score estimation will be inaccurate, which will further lead to the low quality of the 
sampled data~\cite{song2019generative}. If the wave signal is contaminated with a very small scale noise, then the contaminated  
wave signal will spread to the 
entire space $\mathbb{R}^d$ instead of being limited to a small low-dimensional manifold. When using perturbed wave signal as input,
 the following denoising score matching (DSM) loss~\cite{vincent2011connection,song2019generative} 
\begin{align}
  &\text{DSM loss}=\mathbb{E}_{t\sim [0,T]}\mathbb{E}_{\mathbf{x}(0)\sim p_{mel}(\mathbf{x}(0))}  \mathbb{E}_{\mathbf{x}(t)\sim p(\mathbf{x}(t)|\mathbf{x}(0))} 
  \nonumber \\& [ \frac{1}{2}\parallel \mathfrak{S}_{\theta}(\mathbf{x}(t),t,\mathbf{m}) -\nabla_{\mathbf{x}(t)}  \log  p(\mathbf{x}(t)|\mathbf{x}(0))\parallel^2 ].
  \label{eq:dsm_loss}
\end{align}
is equal to the loss~(\ref{eq:esm_loss}) of a non-parametric (e.g. Parzen windows densiy) estimator~\cite{vincent2011connection}.
This DSM loss is used in this paper to train the score prediction network.
If we can accurately estimate the score $\nabla_{\mathbf{x}(t)}  \log  p(\mathbf{x}(t))$ of the distribution, then we can generate 
 wave sample data from the original distribution.  

It should be \textbf{noted} that in the experiment we found that the choice of training loss is very critical. For  It\^oWave, the $\mathcal{L}2$ loss is more appropriate than $\mathcal{L}1$ loss. 

Generally the transition densities $p(\mathbf{x}(t)|\mathbf{x}(0))$ and the score 
$\nabla_{\mathbf{x}(t)}  \log  p(\mathbf{x}(t)|\mathbf{x}(0))$ 
in the DSM loss are difficult to calculate, but for linear SDE, these values 
have close formulas~\cite{sarkka2019applied}.

\subsection{Linear SDE and transition densities}
\label{sec:linear_sde}

Empirically it is found that different types of linear SDE for different audio generation tasks, e.g. the 
variance exploding (VE) SDE~\cite{song2020score} is much suitable for wave generation.
VE SDE is of the following form
 \begin{align}
  \left\{
  \begin{array}{lr}
 d\mathbf{X}=\sigma_0(\frac{\sigma_1}{\sigma_0})^t\sqrt{2\log \frac{\sigma_1}{\sigma_0}}d\mathbf{W} &  \\
 \mathbf{X}(0)=\mathbf{x}(0)\sim \int p_{mel}(\mathbf{x})\mathcal{N}(\mathbf{x}(0);
 \mathbf{x},\sigma_0^2\mathbf{I})d\mathbf{x}, &
\end{array}
\right.
 \label{eq:ve_sde}
\end{align}
where $\sigma_0=0.01 < \sigma_1$.

Then the differential equation satisfied by the mean and variance of the transition densities $p(\mathbf{x}(t)|\mathbf{x}(0))$ is as follows
\begin{align}
  \left\{
  \begin{array}{lr}
    \frac{d \mathbf{m}(t)}{dt} = \mathbf{0} &  \\
    \frac{d \mathbf{V}(t)}{dt} = 2\sigma^2_0(\frac{\sigma_1}{\sigma_0})^{2t}\log \frac{\sigma_1}{\sigma_0}\mathbf{I}.&
\end{array}
\right.
  \label{eq:mean_var_pde_no1}
 \end{align}
 Solving the above equation, and choose $\sigma_1$ makes $2\log \frac{\sigma_1}{\sigma_0}=1$, we get
the transition density of this as
\begin{align}
&\quad p(\mathbf{x}(t)|\mathbf{x}(0))\nonumber \\ &=\mathcal{N}\left(\mathbf{x}(t);
\mathbf{x}(0),\left[\sigma_0^2(\frac{\sigma_1}{\sigma_0})^{2t}-\sigma_0^2\right]\mathbf{I}\right).
 \label{eq:ve_sde_trans_density}
\end{align}

The score of the VE linear SDE is
\begin{align}
  &\quad \nabla_{\mathbf{x}(t)}\log p(\mathbf{x}(t)|\mathbf{x}(0)) \nonumber \\&=\nabla_{\mathbf{x}(t)} \log \mathcal{N}\left(\mathbf{x}(t); 
  \mathbf{x}(0),\left[\sigma_0^2(\frac{\sigma_1}{\sigma_0})^{2t}-\sigma_0^2\right]\mathbf{I}\right)  \nonumber \\&
  =\nabla_{\mathbf{x}(t)} [ -\frac{d}{2}\log \left[2\pi \left(\sigma_0^2(\frac{\sigma_1}{\sigma_0})^{2t}-\sigma_0^2\right)\right]\nonumber \\& - 
  \frac{\parallel\mathbf{x}(t) - \mathbf{x}(0)\parallel^2}{2 \left( \sigma_0^2(\frac{\sigma_1}{\sigma_0})^{2t}-\sigma_0^2  \right)} ]
  = -  \frac{\mathbf{x}(t) - \mathbf{x}(0)}{ \sigma_0^2(\frac{\sigma_1}{\sigma_0})^{2t}-\sigma_0^2}. 
   \label{eq:ve_sde_score}
  \end{align}

The prior distribution $p(\mathbf{x}(T))$ is a  Gaussian
\begin{align}
\mathcal{N}\left(\mathbf{x}(T);\mathbf{0},\sigma_1^2\mathbf{I}\right)=
\frac{\exp (-\frac{1}{2\sigma_1^2}\left\lVert \mathbf{x}(T)\right\rVert^2) }{\sigma_1^d\sqrt{(2\pi)^d}},
\label{eq:ve_sde_prior}
\end{align}
thus $\log p(\mathbf{x}(T)) = -\frac{d}{2}\log(2\pi\sigma_1^2)
  - \frac{1}{2\sigma_1^2}\left\lVert \mathbf{x}(T)\right\rVert^2 $.

\subsection{Training and wave sampling algorithms}
\label{sec:training_algorithm}

Based on subsections~\ref{sec:dist_transormation_sde} and~\ref{sec:score_estimation}, we can get the training procedure of the score networks based 
on general SDEs, as shown in Algorithm~\ref{alg:mel_sde_score_training}. 

\begin{algorithm}[!h]
  \caption{Training of the score network in VE SDE-based wave generation model}
  \label{alg:mel_sde_score_training}
  
  \textbf{Input and initialization}: The wave $\mathbf{x}$  and the corresponding mel spectrogram condition $\mathbf{m}$, the diffusion time $T$.
  
  1: \textbf{for} $k=0, 1, \cdots$
  
  2:  \quad  Uniformly sample $t$ from $[0, T]$.

  3:  \quad \quad Randomly sample batch of $\mathbf{x}$ and  $\mathbf{m}$, let $\mathbf{x}(0)=\mathbf{x}$. Sample $\mathbf{x}(t)$ from the 
          distribution $\mathcal{N}\left(\mathbf{x}(t);
          \mathbf{x}(0),\left[\sigma_0^2(\frac{\sigma_1}{\sigma_0})^{2t}-\sigma_0^2\right]\mathbf{I}\right)$, compute the target score 
          as $-  \frac{\mathbf{x}(t) - \mathbf{x}(0)}{ \sigma_0^2(\frac{\sigma_1}{\sigma_0})^{2t}-\sigma_0^2 } $. Average the following
          \begin{align}
           \text{DSM } loss=
          \parallel \mathfrak{S}_{\theta_k}(\mathbf{x}(t),t,\mathbf{m}) +  \frac{\mathbf{x}(t) - \mathbf{x}(0)}{ \sigma_0^2(\frac{\sigma_1}{\sigma_0})^{2t}-\sigma_0^2 }\parallel_1 . \nonumber
         \end{align} 
  
4:  \quad \quad Do the back-propagation and the parameter updating of $\mathfrak{S}_{\theta}$.

5: \quad $k \leftarrow k+1$.

 6: \textbf{Until} stopping conditions are satisfied and $\mathfrak{S}_{\theta_k}$ converges, e.g. to $\mathfrak{S}_{\theta_*}$.
  
  \textbf{Output}: $\mathfrak{S}_{\theta_*}$.
  \end{algorithm}

After we get the optimal score network  $\mathfrak{S}_{\theta_*}$ through loss minimization, thus we can get the gradient of log value of the distribution 
probability density of the  wave with $\mathfrak{S}_{\theta_*}(\mathbf{x}(t),t,\mathbf{m})$.
Then we can use Langevin dynamics or the reverse-time It\^o SDE~(\ref{eq:reverse_ito_sde})
to generate the wave corresponding to the specific mel spectrogram $\mathbf{m}$.
The reverse-time SDE~(\ref{eq:reverse_ito_sde}) 
can be solved and used to generate target audio data.
Assuming that the time schedule is fixed, the discretization of the diffusion process~(\ref{eq:ito_sde}) is as follows
\begin{align}
  \left\{
  \begin{array}{lr}
 \quad \mathbf{X}(i\Delta t+\Delta t) - \mathbf{X}(i\Delta t)&  \\ =\mathbf{f}(\mathbf{X}(i\Delta t),i\Delta t)\Delta t+g(i\Delta t)\mathbf{\xi}(i\Delta t)&  \\  (i=0,1,\cdots, N-1)&  \\
 \mathbf{X}(0)=\mathbf{x}(0) &
\end{array}
\right.
 \label{eq:discret_ito_sde}
\end{align}
since $d\mathbf{W}$ is a wide sense stationary white noise process~\cite{evans2012introduction}, which is denoted as $\mathbf{\xi}(\cdot)\sim 
\mathcal{N}\left(\mathbf{0},\mathbf{I}\right)$ in this paper.

The corresponding discretization of the reverse-time diffusion process~(\ref{eq:reverse_ito_sde}) is 
\begin{align}
  \left\{
  \begin{array}{lr}
     \quad \mathbf{X}(i\Delta t) - \mathbf{X}(i\Delta t+\Delta t) &\\=\mathbf{f}(\mathbf{X}(i\Delta t+\Delta t),i\Delta t+\Delta t)  (-\Delta t) &  \\
    -g(i\Delta t+\Delta t)^2 \mathfrak{S}_{\theta_*}(\mathbf{X}(i\Delta t+\Delta t),i\Delta t+\Delta t, \textbf{m}) (-\Delta t)&  \\ + g(i\Delta t+\Delta t)\mathbf{\xi}(i\Delta t) &  \\
 \mathbf{X}(T)=\mathbf{x}(T)& \\
 T = N\Delta t, \quad i=0,1,\cdots, N-1 &
\end{array}
\right.
 \label{eq:discret_reverse_ito_sde}
\end{align}

In this paper, we use the strategy of~\cite{song2020score}, which means that at each time step,  Langevin dynamics is used to predict first, and then reverse-time 
It\^o SDE~(\ref{eq:discret_reverse_ito_sde}) is used to revises the first predicted result.

The generation algorithm of wave based on VE linear SDE is as follows
\begin{algorithm}[H]
  \caption{VE SDE-based It\^oWave wave generation algorithm}
  \label{alg:general_sde_sampling}
  
  \textbf{Input and initialization}: the score network $\mathfrak{S}_{\theta_*}$, input mel-spectrogram $\mathbf{m}$, and $\mathbf{x}( N\Delta t)\sim 
  \mathcal{N}\left(\mathbf{0},\sigma_1\mathbf{I}\right)$.
  
  1: \textbf{for} $k=N-1, \cdots, 0$
  
  2:  \quad   $\mathbf{x}(k\Delta t) = \mathbf{x}(k\Delta t+\Delta t) 
  + 2\sigma_0^2(\frac{\sigma_1}{\sigma_0})^{2k\Delta t+2\Delta t}\log \frac{\sigma_1}{\sigma_0}
  \mathfrak{S}_{\theta_*}(\mathbf{x}(k\Delta t+\Delta t),k\Delta t+\Delta t, \mathbf{m}) \Delta t + 
  \sigma_0(\frac{\sigma_1}{\sigma_0})^{k\Delta t+\Delta t}\sqrt{2\log \frac{\sigma_1}{\sigma_0}} \mathbf{\xi}(k\Delta t) $

 3: \quad  $\mathbf{x}(k\Delta t) \leftarrow \mathbf{x}(k\Delta t) + \epsilon_k \mathfrak{S}_{\theta_*}
 (\mathbf{x}(k\Delta t),k\Delta t, \mathbf{m})  +\sqrt{2\epsilon_k}\mathbf{\xi}(k\Delta t)$
  
  4: \quad $k \leftarrow k-1$.
  
  \textbf{Output}: The generated wave $\mathbf{x}(0)$.
  \end{algorithm}

\subsection{Architectures of   $\mathfrak{S}_{\theta}(\mathbf{x}(t),t,\mathbf{m})$.}
\label{sec:architecture}

It was found that although the score network model does not have as strict restrictions 
on the network structure as the flow model~\cite{miao2020flow,prenger2019waveglow,valle2020flowtron}, not all network structures are suitable for score prediction.

\begin{figure}[th]
    \centering
    \hspace{-5mm}
    \includegraphics[width=1.0\linewidth]{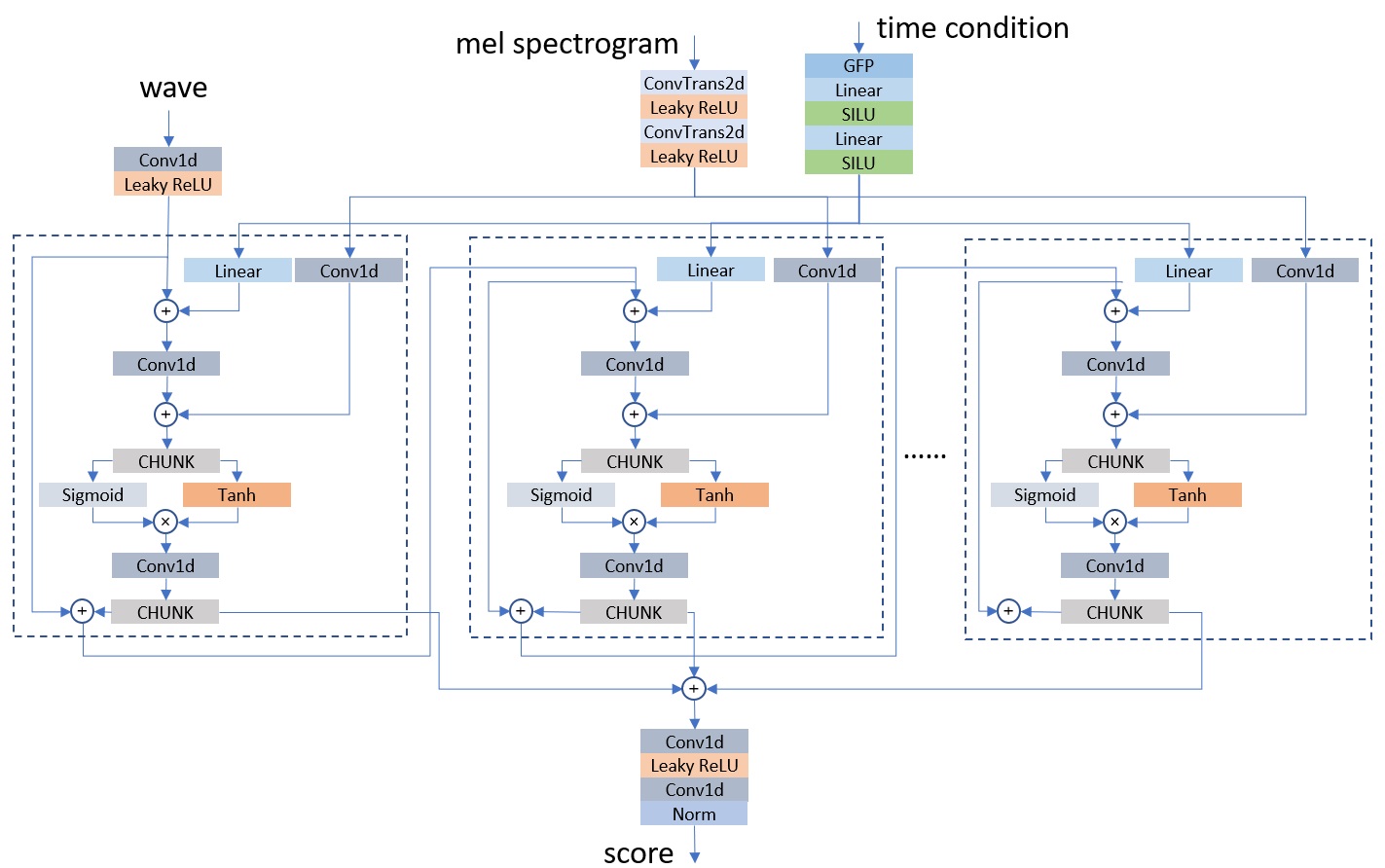}
    \hspace{-5mm}
  \caption{
  The architecture of It\^oWave.
      }
    \label{itowave_arch}
    \end{figure}

The structure of \textbf{It\^oWave}'s score prediction network $\mathfrak{S}_{\theta}(\mathbf{x}(t),t,\mathbf{m})$ is shown in Figure~\ref{itowave_arch}. The input is the wave to be generated, and the conditional 
input has mel spectrogram and time $t$. 
  The output is the score at time $t$. All three types of input require preprocessing processes. The preprocessing of the wave is through a 
  convolution layer; the preprocessing of mel spectrogram is based on the upsampling by two  transposed convolution layers. 
  After all inputs are preprocessed, they will be sent to the 
  most critical module of $\mathfrak{S}_{\theta}(\mathbf{x}(t),t,\mathbf{m})$, which  is of several serially connected dilated residual blocks. The main input 
  of the dilated residual block is the wave, and the step time condition and mel spectrogram condition will be input into these dilated residual blocks 
  one after another, and added to the feature map after the transformation of the wave signal. Similarly, there are two outputs of each dilated 
  residual block, one is the state, which is used for input to the next residual block, and the other is the final output. The 
  advantage of this is the ability to synthesize information of different granularities. Finally, the outputs of all residual 
  blocks are summed and then pass through two convolution layers as the final output score.

\section{Experiments}
\label{sec:experiments}


\subsection{Dataset and setup}

The data set we use is LJSpeech~\cite{ito2017lj}, a single female speech database, with a total of 24 hours, 13100 sentences, randomly divided into 13000/50/50 for 
training/verification/testing. The sampling rate is 22050.  In the experiment, for mel spectrogram, the window length is 1024, hop length is 256, the number of mel channels is 80. 
We use the same Adam~\cite{kingma2014adam} training algorithm for It\^oWave. We have done 
quantitative evaluations based on mean opinion score (MOS) 
with other state-of-the-art methods on  It\^oWave.
We 
compared with WaveNet~\cite{oord2016wavenet}, WaveGlow~\cite{prenger2019waveglow}, 
Diffwave~\cite{kong2020diffwave}, and WaveGrad~\cite{chen2020wavegrad}.
All experiments were performed on a GeForce RTX 3090 GPU with 24G memory.

\subsection{Results and discussion}

In order to verify the naturalness and fidelity of the synthesized voice, 
we randomly select 40  from 50 test data for each subject, 
and then let the subject give the synthesized sound a MOS score of 0-5.

\begin{figure}[!h]
  \centering
  \subfigure[The waveform in the first step.]{
  \begin{minipage}[t]{0.5\linewidth}
  \centering
  \includegraphics[width=1.5in]{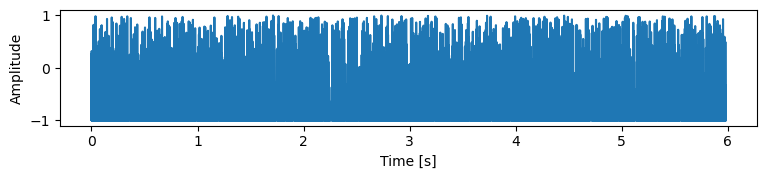}
  \label{tobe_itowave_diffusion_step1}
  \end{minipage}%
  }%
  \subfigure[The waveform at the 200th step.]{
  \begin{minipage}[t]{0.5\linewidth}
  \centering
  \includegraphics[width=1.5in]{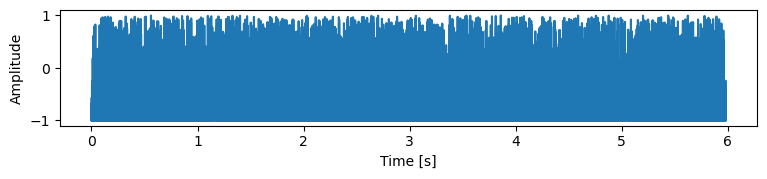}
  \label{tobe_itowave_diffusion_step200}
  \end{minipage}%
  }%

  \subfigure[The waveform at the 300th step.]{
    \begin{minipage}[t]{0.5\linewidth}
    \centering
    \includegraphics[width=1.5in]{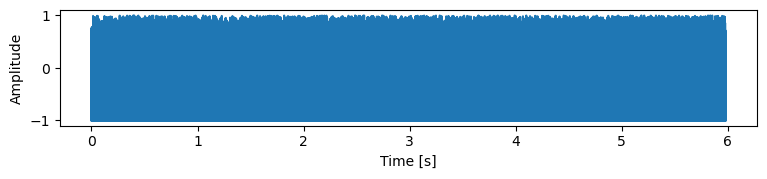}
    \label{tobe_itowave_diffusion_step300}
    \end{minipage}%
    }%
    \subfigure[The waveform at the 400th step.]{
      \begin{minipage}[t]{0.5\linewidth}
      \centering
      \includegraphics[width=1.5in]{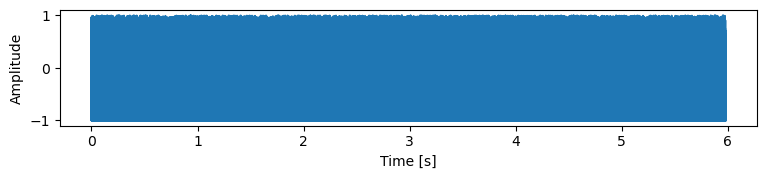}
      \label{tobe_itowave_diffusion_step400}
      \end{minipage}%
      }%

      \subfigure[The waveform at the 500th step.]{
        \begin{minipage}[t]{0.5\linewidth}
        \centering
        \includegraphics[width=1.5in]{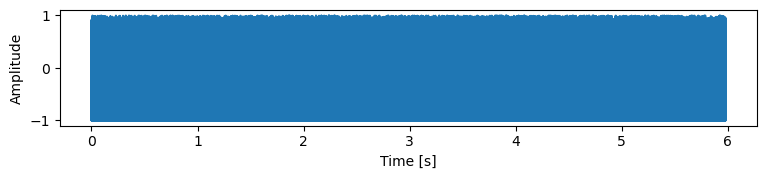}
        \label{tobe_itowave_diffusion_step500}
        \end{minipage}%
        }%
        \subfigure[The waveform at the 600th step.]{
          \begin{minipage}[t]{0.5\linewidth}
          \centering
          \includegraphics[width=1.5in]{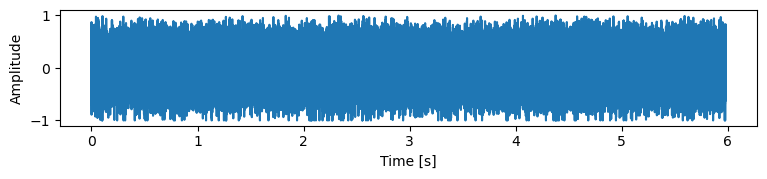}
          \label{tobe_itowave_diffusion_step600}
          \end{minipage}%
          }%

          \subfigure[The waveform at the 700th step.]{
            \begin{minipage}[t]{0.5\linewidth}
            \centering
            \includegraphics[width=1.5in]{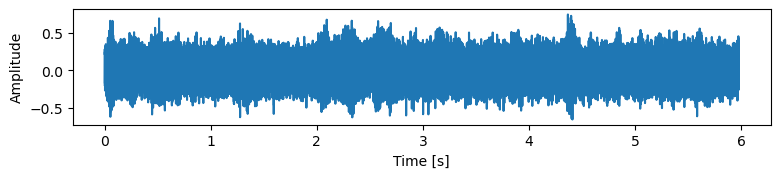}
            \label{tobe_itowave_diffusion_step700}
            \end{minipage}%
            }%
            \subfigure[The waveform at the 800th step.]{
              \begin{minipage}[t]{0.5\linewidth}
              \centering
              \includegraphics[width=1.5in]{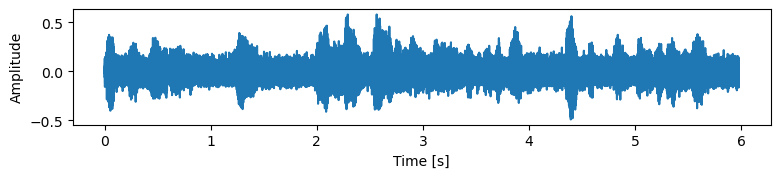}
              \label{tobe_itowave_diffusion_step800}
              \end{minipage}%
              }%

              \subfigure[The waveform at the 900th step.]{
                \begin{minipage}[t]{0.5\linewidth}
                \centering
                \includegraphics[width=1.5in]{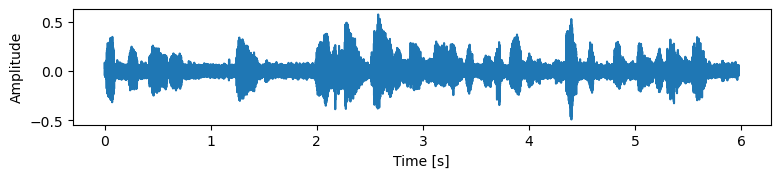}
                \label{tobe_itowave_diffusion_step900}
                \end{minipage}%
                }%
                \subfigure[The waveform at the 1000th step.]{
                  \begin{minipage}[t]{0.5\linewidth}
                  \centering
                  \includegraphics[width=1.5in]{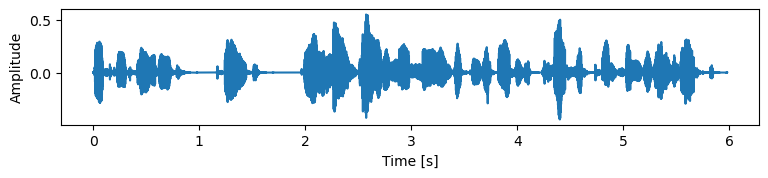}
                  \label{tobe_itowave_diffusion_step1000}
                  \end{minipage}%
                  }%

  \centering
  \caption{Conditioned on the frequency spectrum of the sentence LJ032-0167 in LJSpeech,  It\^oWave generates the corresponding voice step 
  by step from the Gaussian signal. The corresponding text is ``he concluded, quote, there is no doubt in my mind that these fibers 
  could have come from this shirt.''}
 \label{tobe_itowave_diffusion}
  \end{figure}

For It\^oWave,  the original mel spectrogram of the test set was used as the condition input to the score estimation network.
It\^oWave uses 30 residual layers.
For the comparison methods, 

The results are shown in Table~\ref{tab:vocoder}, and you can see that It\^oWave scores the best. 
It has approached the true value of ground truth.  As shown in Figure~\ref{tobe_itowave_diffusion}, we can see how It\^oWave gradually turns white noise into meaningful wave.

\begin{table}[!h]
  \caption[mos_vocoder]{MOS with 95\% confience  in a comparative study of different state-of-the-art vocoders on the 
  test set of LJspeech dataset.}\label{tab:vocoder}
  \centering
  \begin{tabular}{c|c}
  \hline
  \hline
  Methods & MOS\\
  \hline
  \hline
  Ground truth &4.45$\pm$ 0.07\\
  WaveNet &4.3$\pm$  0.130 \\
 WaveGlow &3.95$\pm$  0.161\\
 DiffWave &4.325$\pm$  0.123\\
 WaveGrad &4.1$\pm$  0.158\\
  \hline
  \hline
  It\^oWave  & 4.35$\pm$ 0.115 \\
  \hline
  \end{tabular}
  \end{table}

\section{Conclusion}
\label{sec:conclusion}

This paper proposes  a new vocoder  It\^oWave based on linear It\^o SDE. 
Under conditional input, It\^oWave can continuously transform simple distributions into corresponding  wave 
data through reverse-time linear SDE and Langevin dynamic. It\^oWave 
use neural networks to predict the required score for reverse-time 
linear SDE and Langevin dynamic sampling, which is the gradient of the log probability density at a specific time. 
For It\^oWave, 
we designed the corresponding effective score prediction networks. Experiments show that the MOS of  It\^oWave 
can achieved the state-of-the-art respectively.



\bibliographystyle{IEEEbib}
\bibliography{icassp2022}

\end{document}